\documentclass{aa}  

\usepackage{graphicx}
\usepackage{subfig}
\usepackage{fixltx2e}
\usepackage{siunitx}

\usepackage{txfonts}
\begin{document}

   \title{Direct detection of scattered light gaps in the transitional disk around HD\,97048 with VLT/SPHERE \thanks{Based on data collected at the European Southern Observatory, Chile (ESO Programs 096.C-0248, 096.C-0241, 077.C-0106).}}


   \author{C. Ginski\inst{1}
	  \and T. Stolker\inst{2}
	  \and P. Pinilla\inst{1}
	  \and C. Dominik\inst{2}
	  \and A. Boccaletti\inst{3}
	  \and J. de Boer\inst{1}
	  \and M. Benisty\inst{4,5}
	  \and B. Biller\inst{6}
	  \and M. Feldt\inst{7}
	  \and A. Garufi\inst{8,9}
	  \and C.U. Keller\inst{1}
	  \and M. Kenworthy\inst{1}
	  \and A.L. Maire\inst{7}
	  \and F. M\'{e}nard\inst{4,5}
	  \and D. Mesa \inst{10}
	  \and J. Milli\inst{11}
	  \and M. Min\inst{12,2}
	  \and C. Pinte\inst{4,5}
	  \and S.P. Quanz\inst{8}
	  \and R. van Boekel\inst{7}
	  \and M. Bonnefoy\inst{4,5}
	  \and G. Chauvin\inst{4,5}
	  \and S. Desidera\inst{10}
	  \and R. Gratton\inst{10}
	  \and J. H. V. Girard\inst{11}
	  \and M. Keppler\inst{4,5}
	  \and T. Kopytova\inst{7}
	  \and A.-M. Lagrange\inst{4,5}
	  \and M. Langlois\inst{13,14}
	  \and D. Rouan\inst{3}
	  \and A. Vigan\inst{14}
          }

   \institute{Leiden Observatory, Leiden University, P.O. Box 9513, 2300 RA Leiden, The Netherlands\\
              \email{ginski@strw.leidenuniv.nl}
              \and Anton Pannekoek Institute for Astronomy, University of Amsterdam, Science Park 904, 1098 XH Amsterdam, The Netherlands
              \and LESIA, Observatoire de Paris-Meudon, CNRS, Universit\'{e} Pierre et Marie Curie, Universit\'{e} Paris Didierot, 5 Place Jules Janssen, F-92195 Meudon, France
              \and Universit\'{e} Grenoble Alpes, IPAG, F-38000 Grenoble, France
	      \and CNRS, IPAG, F-38000 Grenoble, France
	      \and Institute for Astronomy, University of Edinburgh, Blackford Hill View, Edinburgh EH9 3HJ, UK 	      
	      \and Max Planck Institute for Astronomy, K\"{o}nigstuhl 17, 69117 Heidelberg, Germany
	      \and Institute for Astronomy, ETH Zurich, Wolfgang-Pauli-Strasse 27, 8093 Zurich, Switzerland
	      \and Universidad Auton\'{o}noma de Madrid, Dpto. F\'{i}sica Te\'{o}rica, M\'{o}dulo 15, Facultad de Ciencias, Campus de Cantoblanco, E-28049 Madrid, Spain
	      \and INAF-Osservatorio Astronomico di Padova, Vicolo dell'Osservatorio 5, Padova, Italy, 35122-I 
	      \and ESO, Alonso de C\'{o}rdova 3107, Vitacura, Casilla 19001, Santiago de Chile, Chile
              \and SRON Netherlands Institute for Space Research, Sorbonnelaan 2, 3584 CA Utrecht, The Netherlands
              \and CRAL, UMR 5574, CNRS, Universit\'{e} Lyon 1, 9 avenue Charles Andr\'{e}, 69561 Saint Genis Laval Cedex, France
	      \and Aix Marseille Universit\'{e}, CNRS, LAM (Laboratoire d'Astrophysique de Marseille) UMR 7326, 13388, Marseille, France
             }

   \date{Received July XX, 2016; accepted XX, 2016}

 
  \abstract
   {}
   {We studied the well known circumstellar disk around the Herbig Ae/Be star HD\,97048 with high angular resolution to reveal undetected structures in the disk, which may be indicative of disk evolutionary processes such as planet formation.}
   {We used the IRDIS near-IR subsystem of the extreme adaptive optics imager SPHERE at the ESO/VLT to study the scattered light from the circumstellar disk via high resolution polarimetry and angular differential imaging.}
   {We imaged the disk in unprecedented detail and revealed four ring-like brightness enhancements and corresponding gaps in the scattered light from the disk surface with radii between 39\,au and 341\,au. We derived the inclination and position angle as well as the height of the scattering surface of the disk from our observational data.
   We found that the surface height profile can be described by a single power law up to a separation $\sim$270\,au. Using the surface height profile we measured the scattering phase function of the disk and found that it is well consistent with theoretical models of compact dust aggregates.
   We discuss the origin of the detected features and find that low mass ($\leq$1\,M$_{\mathrm{Jup}}$) nascent planets are a possible explanation.}
   {}


   \maketitle
%

\section{Introduction}
In the past two decades a large variety of planetary systems have been discovered by indirect observation techniques such as radial velocity surveys (see e.g. \citealt{2007ARA&A..45..397U}) and transit searches (see \citealt{2013ApJS..204...24B}). To understand the formation process of these planets we need to study the initial conditions and evolution of protoplanetary disks. Intermediate mass Herbig Ae/Be stars are prime targets for such studies due to the massive disks they are hosting.  
High resolution imaging observations have been performed for a number of these disks in the past few years, e.g. \cite{2012ApJ...748L..22M} and \cite{2015ApJ...813L...2W} showed resolved spiral arms in the disks around SAO\,206462 and HD\,100453, while \cite{1999A&A...350L..51A}, \cite{2015MNRAS.450.4446B}, \cite{2016ApJ...819L..26C} and \cite{2016A&A...590L...7P} found asymmetric rings and gaps with increasing detail in the transitional disk around HD\,141569\,A. These structures may be signposts of ongoing planet formation.
Indeed signatures of massive accreting protoplanets were directly detected in recent years in the disks of LkCa\,15 (\citealt{2012ApJ...745....5K}, \citealt{2015Natur.527..342S}) and HD\,100546 (\citealt{2013ApJ...766L...1Q}, \citealt{2015ApJ...807...64Q}, \citealt{2015ApJ...814L..27C}). 
Furthermore, several prominent young, massive planets orbiting stars of early spectral type were found through direct imaging observations, such as $\beta$\,Pic\,b (\citealt{2009A&A...493L..21L}), HR\,8799\,b,c,d,e (\citealt{2008Sci...322.1348M}, \citealt{2010Natur.468.1080M}), HD\,95086\,b (\citealt{2013ApJ...772L..15R}), 51\,Eri\,b (\citealt{2015Sci...350...64M}) and most recently HD\,131399\,Ab (\citealt{2016Sci...353..673W}). Thus these stars enable us to study the beginning and the end of the planet formation process at scales that we can spatially resolve.\\
HD\,97048 is a young (2-3\,Myr, \citealt{1998A&A...330..145V}, \citealt{2006Sci...314..621L}), well studied Herbig Ae/Be star at a distance of 158$^{+16}_{-14}$pc (\citealt{2007A&A...474..653V}). The mass of the star is estimated to be 2.5$\pm$0.2\,M$_\odot$ (\citealt{1998A&A...330..145V}).
The star is well known to harbor a large ($\geq$\,600\,au, \citealt{2007AJ....133.2122D}) circumstellar disk. The disk was first spatially resolved in spectroscopy by \cite{2004A&A...418..177V} using TIMMI2 in the mid-IR and in imaging by \cite{2006Sci...314..621L} and \cite{2007A&A...470..625D} using VLT/VISIR with filters sensitive to PAH emission features in the mid-IR. Follow-up studies revealed the extended disk (between 320\,au and 630\,au) in optical scattered light (\citealt{2007AJ....133.2122D}), as well as in near-IR polarized light (\citealt{2012A&A...538A..92Q}). However, no structures were directly detected in the disk. From SED fitting combined with resolved Q-band imaging, \cite{2013A&A...555A..64M} deduced the presence of a gap and a puffed up outer disk inner edge at 34$\pm$4\,au.\\
We used the SPHERE (Spectro-Polarimetric High-contrast Exoplanet REsearch, \citealt{2008SPIE.7014E..18B}) instrument at the ESO-VLT to observe HD\,97048 in polarized and integrated light. By applying the polarized differential imaging technique (PDI, \citealt{2001ApJ...553L.189K}, \citealt{2004IAUS..221..307A}) as well as the angular differential imaging technique (ADI, \citealt{2006ApJ...641..556M}) we were able to remove the flux from the central star to study the circumstellar disk in unprecedented detail. 
In addition, we re-analyzed archival polarimetric VLT/NACO (\citealt{2003SPIE.4841..944L}, \citealt{2003SPIE.4839..140R}) data to compare them to our new observations. We identified four ring-like brightness enhancements and associated brightness decreases that we discuss in the following sections. 


\section{Observation and data reduction}

\subsection{IRDIS polarimetric observations}

The HD\,97048 system was observed as part of the ongoing SPHERE guaranteed time program to search and characterize circumstellar disks. 
Observations were carried out with the infrared subsystem IRDIS (Infra-Red Dual Imaging and Spectrograph, \citealt{2008SPIE.7014E..3LD}) in the dual polarization imaging mode (DPI, \citealt{2014SPIE.9147E..1RL}) on February 20th 2016.
Observing conditions were poor during the whole imaging sequence with average seeing above 1.3\,arcsec (coherence time of 3\,ms). We reached a mean Strehl ratio of 45\%. 
We took 31 polarimetric cycles, each consisting of 4 data cubes, one per half wave plate (HWP) position. Each data cube consisted of 2 individual exposures with exposure times of 16\,s, giving a total integration time of 64\,min. All observations were conducted with the broad-band J filter of SPHERE and with the primary star placed behind an apodized Lyot coronagraph with a diameter of 145\,mas.
To determine the accurate position of the star behind the coronagraph, dedicated center calibration frames with 4 satellite spots produced by a waffle pattern applied to the deformable mirror were taken at the beginning and at the end of the observing sequence.
We also took flux calibration images at the beginning and at the end of the sequence with the star moved away from the coronagraph and a neutral density filter inserted to prevent saturation.\\
For data reduction we first applied standard calibrations to each individual image, which consisted of dark subtraction, flat fielding and bad-pixel masking. 
We then averaged the two images in each data cube to obtain one image per HWP position and polarimetric cycle. 
Each exposure contains two orthogonal polarization directions that were recorded simultaneously. We split these images into two individual frames (left and right side, corresponding to parallel and perpendicular polarized beam).
To obtain the individual Q$^+$, Q$^-$, U$^+$ and U$^-$ frames for each polarimetric cycle (corresponding to HWP positions of 0$^\circ$, 45$^\circ$, 22.5$^\circ$ and 67.5$^\circ$), we measured the precise position of the central star using the center calibration frames on the left and right side of the image. 
We then aligned and subtracted the right side of the image from the left side\footnote{We only interpolate one of the two beams to align the two polarization directions to minimize interpolation artifacts.}. To correct for the instrumental polarization introduced by the telescope optics, we then subtracted Q$^-$ from Q$^+$ to obtain a clean Stokes Q image (analog for Stokes U).
We finally averaged all Stokes Q and U images. \\
Since the HWP is only sensitive to instrumental polarization upstream of its position in the optical path, there may be a small amount of instrumental polarization present in the final Q and U images.
This instrumental polarization is thought to be proportional to the total intensity image (Stokes I) as discussed by \cite{2011A&A...531A.102C}. We thus subtracted a scaled version of Stokes I from our Q and U images. 
To obtain the scaling factor we converted our Q and U images to the radial Stokes components Q$_\phi$ and U$_\phi$ by the formulas given in \cite{2006A&A...452..657S}.  
A positive Q$_\phi$ signal corresponds to an azimuthal polarization direction and negative signal corresponds to a radial polarization direction.
U$_\phi$ contains all signal with polarization vector angles that are 45\,deg from an azimuthal or radial orientation. This signal is expected to be small for a centrally illuminated symmetrical disk (\citealt{2015A&A...582L...7C}).
We then determined the scaling factor for our instrumental polarization correction such that the flux in an annulus with an inner radius of 10\,pixel ($\sim$0.12\,arcsec) and an outer radius of 30\,pixel ($\sim$0.36\,arcsec) around the stellar position in the U$_\phi$ frame is minimized.
After this step we created the final Q$_\phi$ and U$_\phi$ images. These final reduced images are shown in Fig.~\ref{hd97048-main}.\\
All images were astrometrically calibrated by observations of the close quadruple system $\Theta^1$\,Ori\,B observed with IRDIS in imaging mode using the same filter and coronagraph as in our polarimetric observations.
Details of the astrometric calibration procedure can be found in \cite{2016A&A...587A..56M}. 
We are using a pixel scale of 12.263$\pm$0.008\,mas/pixel and a true north direction of -1.81$\pm$0.30\,deg. We correct for the anamorphism between the image x and y directions by 
stretching the y direction with a factor of 1.0060$\pm$0.0002.

\subsection{IRDIS angular differential imaging observations}

HD\,97048 has also been observed on March 28th, 2016 as part of the
SHINE (SpHere INfrared survey for Exoplanets) program with IRDIS in Dual Band Imaging (DBI, \citealt{2010MNRAS.407...71V})
mode operating in the H2 - H3 filters ($\lambda_\mathrm{H2} = 1.593 \mu m$,
$\lambda_\mathrm{H3} =1.667 \mu m$). The sky was clear with an average seeing below 1.0\,arcsec (average coherence time of 3\,ms). We obtained 80 frames of 64s each during
which the field of view rotated by 24.5$^{\circ}$. The mean Strehl ratio during the sequence was 70\,\%. The starlight was
blocked with an apodized Lyot coronagraph with a mask diameter of 185\,mas. Out of mask images for photometric calibration purposes as
well as frames with the 4 sattelite spots inserted for astrometric calibration were obtained in the same
sequence.\\
The data were first processed with the Data Reduction Handling (DRH,
\citealt{2008ASPC..394..581P}) tools at the SPHERE Data Center and follow the same reduction as the DPI data, i.e. sky subtraction, flat
fielding, bad pixel removal, and recentering to a common center. The
true north correction is also identical to the DPI data. The two spectral
channels were co-added to achieve a higher signal to noise ratio. \\
Finally we employed three different angular differential imaging routines to suppress the starlight. The results are shown in Fig.~\ref{hd97048-main}.
The simple ADI routine creates a reference image by median-combining all images in the sequence. 
This reference image is then subtracted from each individual frame and they are subsequently de-rotated and combined. 
As a second approach we used a principal component analysis (PCA) to create an orthogonal basis set of the stellar PSF. 
For this purpose we utilized the PynPoint routine by \cite{2012MNRAS.427..948A} with a set of 5 basis components. Note that PynPoint resamples the input images to decrease the pixel size by a factor of 2.
Finally we used the SHINE tool, SpeCal (Galicher et al., in prep.) to suppress the starlight using the TLOCI algorithm (\citealt{2011aoel.confP..25G}, \citealt{2014IAUS..299...48M}). 
TLOCI is a variant of the LOCI algorithm (\citealt{2007ApJ...660..770L}) which defines a linear combination of reference frames to be subtracted in particular areas of the image called subtraction zones. 
It slightly differs from LOCI in the shape of the areas since the so-called optimization zone (used to define the coefficients of the linear combination) is centered on the subtraction zone (where the starlight is actually subtracted). 
Here we used circular but relatively narrow areas of 1.5 fwhm width in radius. The optimization and the subtraction areas are separated by a 1 fwhm (full width at half maximum of the PSF) gap which is neither used for optimization nor  subtraction. 
The selection of reference frames in the temporal sequence is determined by a parameter which limit the amount of subtraction on real field object at any separation (here we set this limit to 20\%). 
Once a TLOCI solution is calculated for each science frame, all frames are de-rotated to a common direction and median-combined. Since the TLOCI processing resulted in the highest signal-to-noise detection of the disk features, the TLOCI image was used for all subsequent analysis. 
However, we recover the same structures with all three processing methods.

\subsection{NACO polarimetric archival data}
\label{naco-red}

\begin{figure*}
\centering
\includegraphics[width=\textwidth]{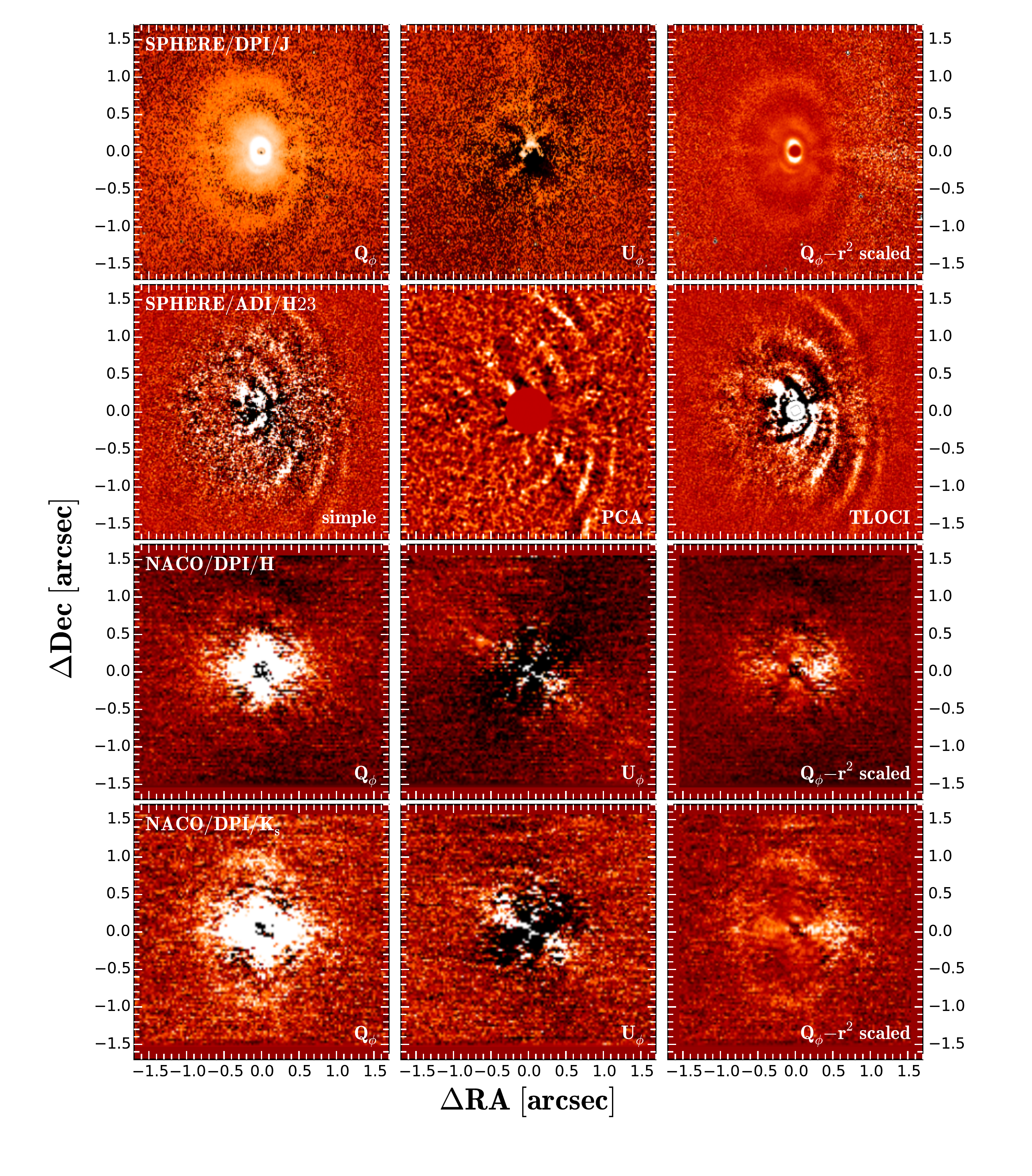}
\caption[]{\textit{1st row: Left and Middle:} Reduced SPHERE DPI Q$_\phi$ and U$_\phi$ images. Color scale and stretch are the same for both images. North is up and East to the left. We have some residual signal in U$_\phi$ close to the center of the image, which may be explained by imperfect centering of the coronagraphic data or by multiple scattering in the inner disk. \textit{1st row: Right:} Q$_\phi$ scaled with the square of the separation from the central star to account for the r$^2$ dependency of the scattered light flux (see Sect. \ref{3.3}). 
\textit{2nd Row:} SPHERE ADI images of the system reduced with 3 different algorithms (simple ADI, PCA and TLOCI). In all cases the H2 and H3 band images were combined to increase the signal.
\textit{3rd and 4th Row: Left and Middle:} NACO H and K$_{\mathrm{s}}$-band Q$_\phi$ and U$_\phi$ images re-reduced by our team. Ring\,2 and Gap\,2 are clearly detected in K$_{\mathrm{s}}$-band. \textit{3rd and 4th Row: Right:} r$^2$-scaled Q$_\phi$ NACO images analog to the corresponding SPHERE image.}
\label{hd97048-main}
\end{figure*}

HD\,97048 was previously observed with polarimetric differential imaging using VLT/NACO. The observations were carried out on April 8th, 2006 and were previously analyzed by \cite{2012A&A...538A..92Q}. 
The system was observed in H and K$_{\mathrm{s}}$-band with a Wollaston prism in place to split the ordinary and extraordinary polarized beams on the detector. A stripe mask was introduced to block light from areas where both beams are overlapping.
In both filters the system was observed with a single frame exposure time of 0.35\,s. For each of the 4 HWP positions, 85 individual integrations were combined, leading to a total integration time of 119\,s per polarimetric cycle. 
In H-band, 12 polarimetric cycles were recorded and 8 in K$_{\mathrm{s}}$-band. Further details can be found in \cite{2012A&A...538A..92Q}. \\
To compare the NACO data to our new SPHERE DPI observations we re-reduced the data set using the same pipeline as for the IRDIS DPI observations, with small adjustments to accommodate the different data structure and 
observation strategy. The main difference in data reduction compared to the IRDIS data set is the absence of a coronagraph during the NACO observations. 
Due to the short single frame exposure time of the NACO data the central star was not saturated\footnote{Some of the frames were outside the linear response regime of the detector, this effect was stronger in K$_{\mathrm{s}}$-band than in H-band. For details we refer to \cite{2012A&A...538A..92Q}.}. We could thus improve the accuracy of the polarimetric differential imaging by re-centering each individual frame.
This was done by fitting a Moffat function to the stellar PSF. The resulting Q$_\phi$ and U$_\phi$ frames are shown in Fig.~\ref{hd97048-main}. 
No dedicated astrometric calibrators were imaged for the VLT/NACO data set. From Fig.~1 in \cite{2014MNRAS.444.2280G} we estimate a true north correction of 0.7$\pm$0.2\,deg. We used the nominal pixel scale provided in the image header.

\section{Disk features and geometry}
\subsection{Description of detected features and comparison with earlier observations}

\begin{figure*}
\centering
\subfloat[][]{
\includegraphics[scale=0.32]{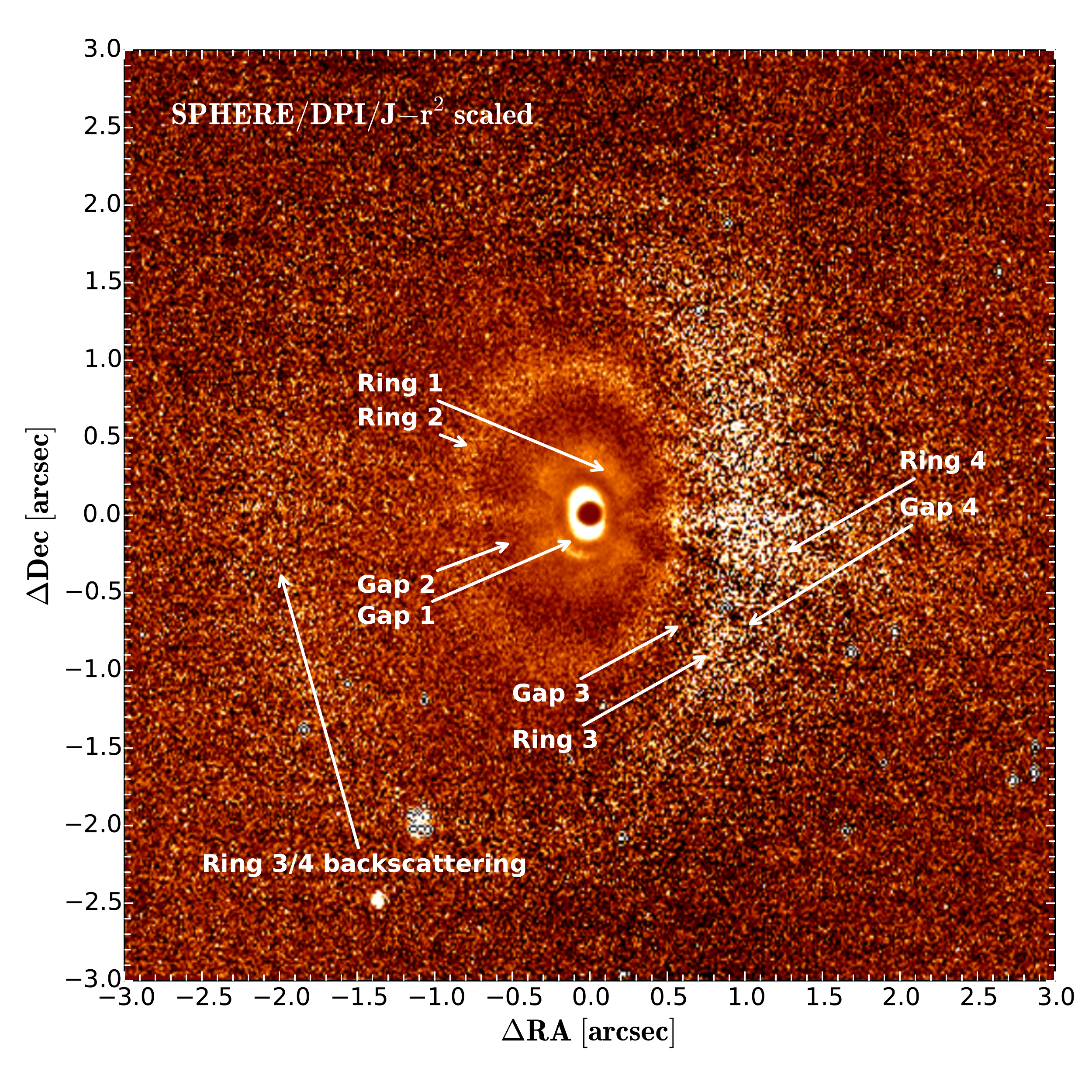}
\label{dpi_anno}
}
\subfloat[][]{
\includegraphics[scale=0.32]{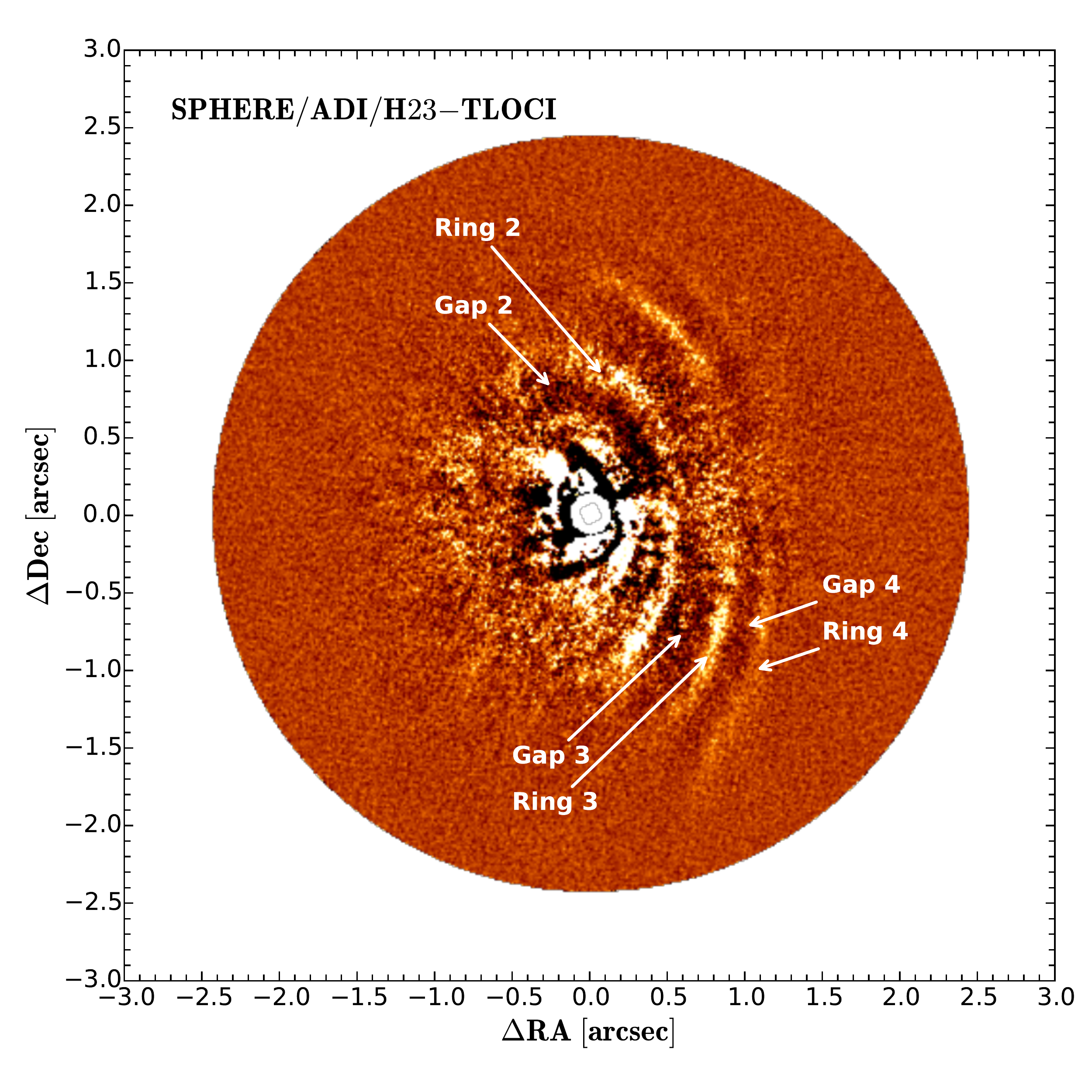}
\label{adi_anno}
}
\caption[]{SPHERE/IRDIS DPI (left) and ADI (right) images of HD\,97048. The recovered disk structure is indicated. In the r$^2$-scaled DPI image we see signal at the (forward scattering) positions of the two outermost rings recovered in the ADI image. 
In addition, we see some signal on the far side of the disk that likely originates from the backscattering side of these outermost rings.} 
\label{ellipse_anno} 
\end{figure*}

\begin{figure}
\centering
\includegraphics[scale=0.5]{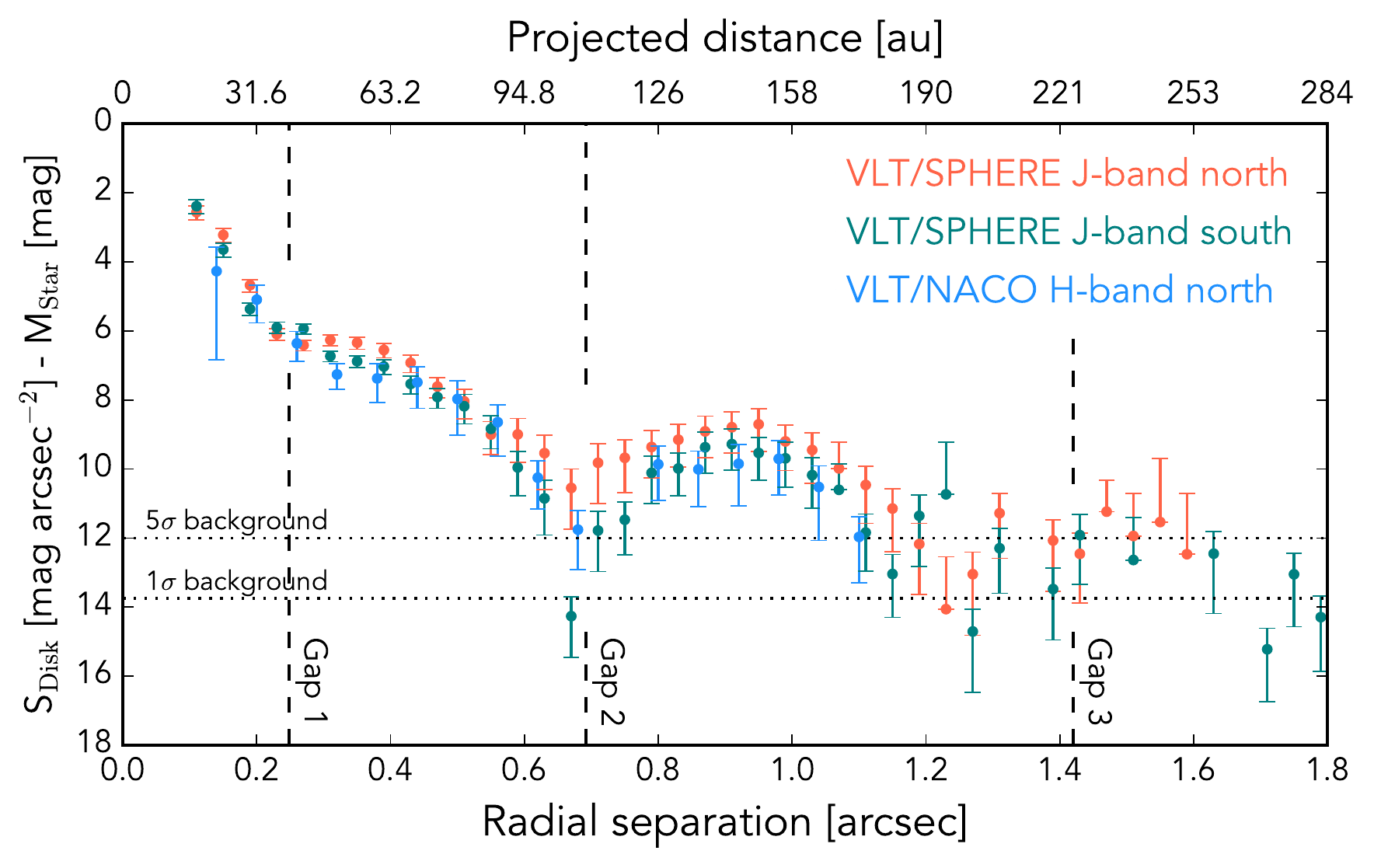}

\caption[]{Calibrated surface brightness profile along the major axis of the disk for our polarimetric SPHERE J-band data, as well as archival NACO H-band data. No significant color effects are visible between the two bands outside of Gap\,1.
The dotted horizontal lines give the sky background limit of our SPHERE data as measured in the Q$_{\mathrm{\phi}}$ image.} 
\label{radialcut}
\end{figure}

We detected several distinct features in our new SPHERE images, which we highlight in Fig~\ref{ellipse_anno}. In polarized light (Fig~\ref{dpi_anno}) we find a continuous disk down to an angular separation of $\sim$75\,mas, i.e. the edge of our coronagraphic mask (the cavity visible further in is only caused by the coronagraph). 
Going along the minor axis towards the West we see a first depression in scattered light at $\sim$0.19\,arcsec (Gap\,1) directly followed by a bright ring (Ring\,1). Ring\,1 is then followed by a continuous disk until $\sim$0.44\,arcsec where we find a second depression in scattered light (Gap\,2), followed again by a ring (Ring\,2). 
In our polarimetric data we then tentatively detect a further gap at $\sim$0.78\,arcsec (Gap\,3) followed by a narrow arc, which we believe is the forward scattering peak of yet another ring (Ring\,3). We conclude that due to the visible compression of the features (specifically Ring\,2) and the slightly increased surface brightness in the western direction that this is the Earth-facing, forward scattering side of the disk.
In our r$^2$-scaled DPI image (see Sect. \ref{3.3}) we also detect some flux on the far (eastern) side of the disk, which we interpret as the extended backscattering signal of Ring\,3 and possibly further out structures. 
Lastly we note that the narrow horizontal bar that is visible in the DPI images is a PSF artifact that was not completely suppressed during data reduction and has no astrophysical origin.\\
In the ADI data (Fig~\ref{adi_anno}), the innermost region of the disk is not accessible due to unavoidable self-subtraction effects in the data reduction process, thus Gap\,1 and Ring\,1 are not (or are only marginally) visible. 
However, we clearly detect the near, forward scattering side of Ring\,2. In the ADI data we now see at high signal to noise that Ring\,2 is indeed followed by a gap (Gap\,3) and another ring (Ring\,3). 
Furthermore we see that beyond Ring\,3 there is yet another gap in the disk at $\sim$1.09\,arcsec (Gap\,4) which is followed by another narrow ring (Ring\,4) of which we again only see the forward scattering side. 
The backscattering far sides of the individual rings are not detected in the ADI image, presumably because their projected sizes are more extended and they are weaker than the forward scattering sides and are thus completely self subtracted.
We note that in the ADI image the rings appear broken along the minor axis, somewhat reminiscent of the recent SPHERE ADI images of the disk around HD\,141569\,A (\citealt{2016A&A...590L...7P}). 
However, due to our complementary DPI data, we clearly see that this is only an ADI artifact in the case of HD\,97048. For a detailed discussion of the effect of the ADI technique on images of circumstellar disks we refer to \cite{2012A&A...545A.111M}.\\
HD\,97048 was previously observed by \cite{2012A&A...538A..92Q} in polarized scattered light in H and K$_{\mathrm{s}}$-band using VLT/NACO. They did not detect any of the features seen in our observations. 
To better compare the NACO data to our new SPHERE observations, we re-reduced the archival data as described in Sec.~\ref{naco-red}. 
While we clearly recover the same cross-shaped inner region of the disk that was also found by \cite{2012A&A...538A..92Q}, we now also marginally detect Ring\,2 in H-band (especially the northern ansa, see Fig.~\ref{hd97048-main}). 
Furthermore we unmistakably detect Ring\,2 in the K$_{\mathrm{s}}$-band image. It is still unclear what produced the strong horizontal signal in the inner part of the disk in the NACO data. 
However, it is not seen in the much higher quality SPHERE data, so it is presumably a PSF or instrumental polarization artifact.\\
We used the NACO data to compare the recovered surface brightness in H band with our measurement in J band to determine if there are any apparent color effects present in the disk. We used H-band only since in K$_{\mathrm{s}}$-band most frames are outside of the linear response regime of the detector close to the stellar position and can thus not be calibrated photometrically. Figure \ref{radialcut} shows the surface brightness profile along the major axis of the disk in northern (PA = $3\degr$, see Sec.~\ref{3.2}) and southern (PA = $183\degr$) direction obtained from the unscaled Q$_\phi$ J-band image. The profile is obtained by selecting all pixels that are azimuthally less than $10\degr$ away from the major axis. 
These pixels are binned radialy in 40\,mas and 60\,mas wide bins for SPHERE and NACO respectively, between 0.05\,arcsec and 2\,arcsec. The error bars show 1$\sigma$ uncertainties of the surface brightness in the unscaled U$_\phi$ image. 
More specifically, we defined annuli at the same radial distance and with the same radial width as the Q$_\phi$ surface brightness bins and determined the standard deviation of the U$_\phi$ pixels within each annulus. 
The profiles in figure \ref{radialcut} show the surface brightness contrast of the polarized scattered light from the disk with the stellar magnitude (assuming 6.67$\pm$0.05\,mag in H-band and 7.27$\pm$0.02\,mag in J-band, \citealt{2003yCat.2246....0C}) which allows us to compare our photometrically calibrated disk brightness in J and H-band.\\ 
We find that the surface brightness in H-band corresponds well with our own J band measurement, i.e. we do not find a strong color dependency of the scattered light surface brightness. 
This could be an indication that the scattered light signal is dominated by particles larger than the wavelength, and for which thus no color dependency would be expected.
However, the disk structures are only detected at very low signal to noise in the NACO image, thus further SPHERE observations in the visible and near infrared wavelength range will be needed to confirm this finding.  

\subsection{Fitting of disk features and scattering height determination}
\label{3.2}

\begin{figure*}
\centering
\subfloat[][]{
\includegraphics[scale=0.32]{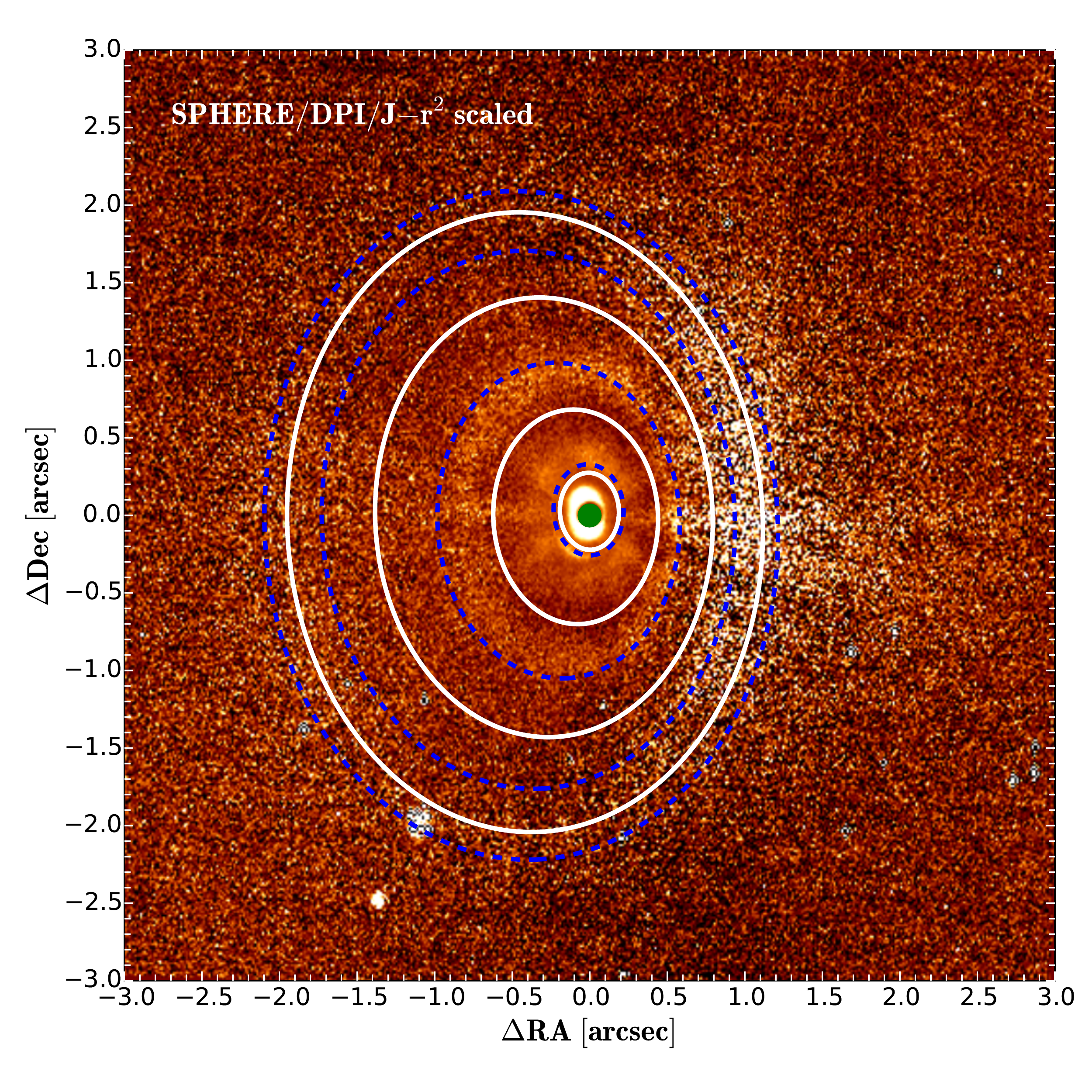}
\label{adi_fits}
}
\subfloat[][]{
\includegraphics[scale=0.32]{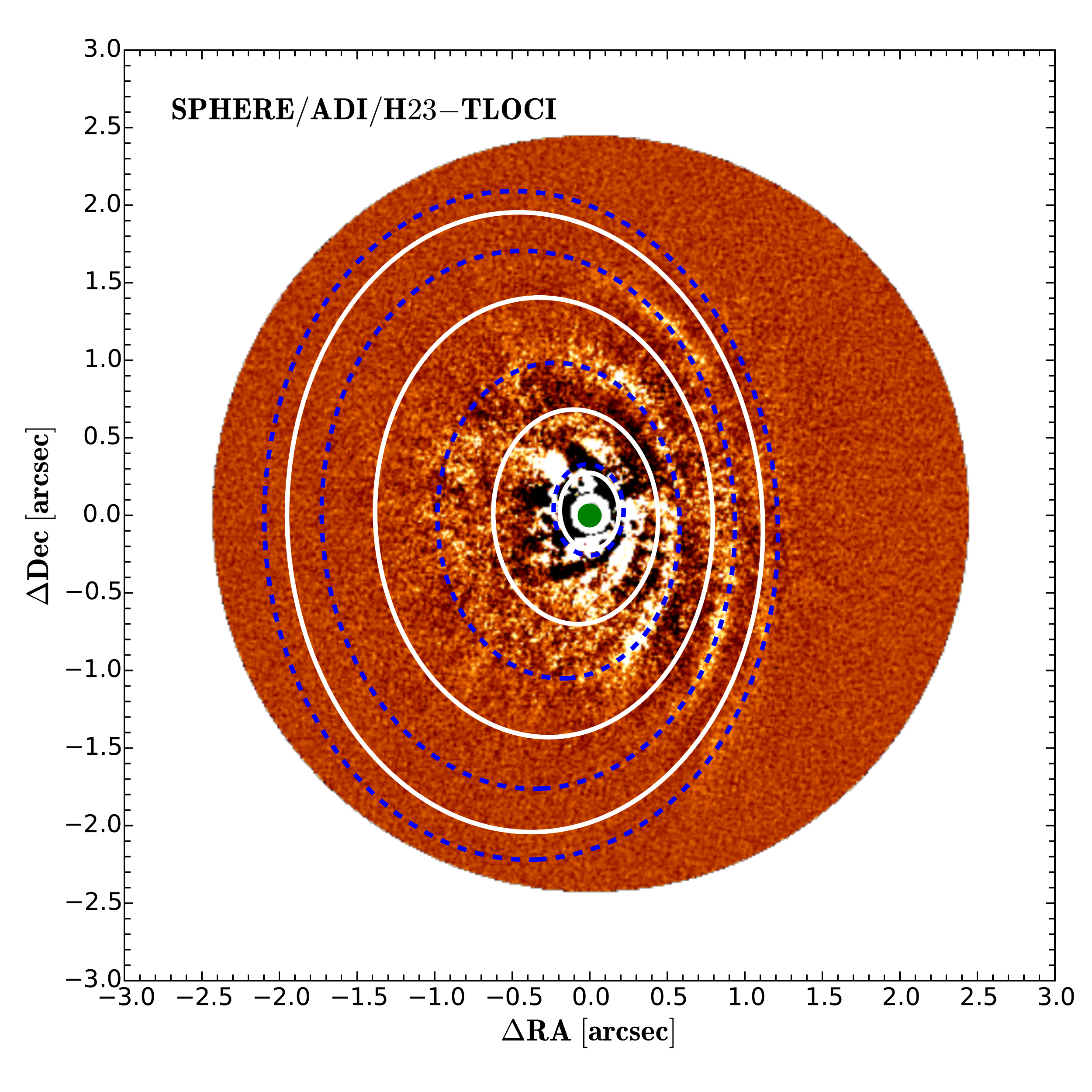}
\label{dpi_fits}
}
\caption[]{SPHERE/IRDIS DPI (left) and ADI (right) images with our best fitting ellipses superimposed. Fitted rings are displayed as blue dashed lines, while fitted gaps are displayed as white solid lines (a color version of this figure is available in the online version of the journal).
The green disk in the center of the system indicates the size of the utilized coronagraphic mask.} 
\label{ellipse_fits} 
\end{figure*}

\begin{figure*}
\centering
\includegraphics[width=\textwidth]{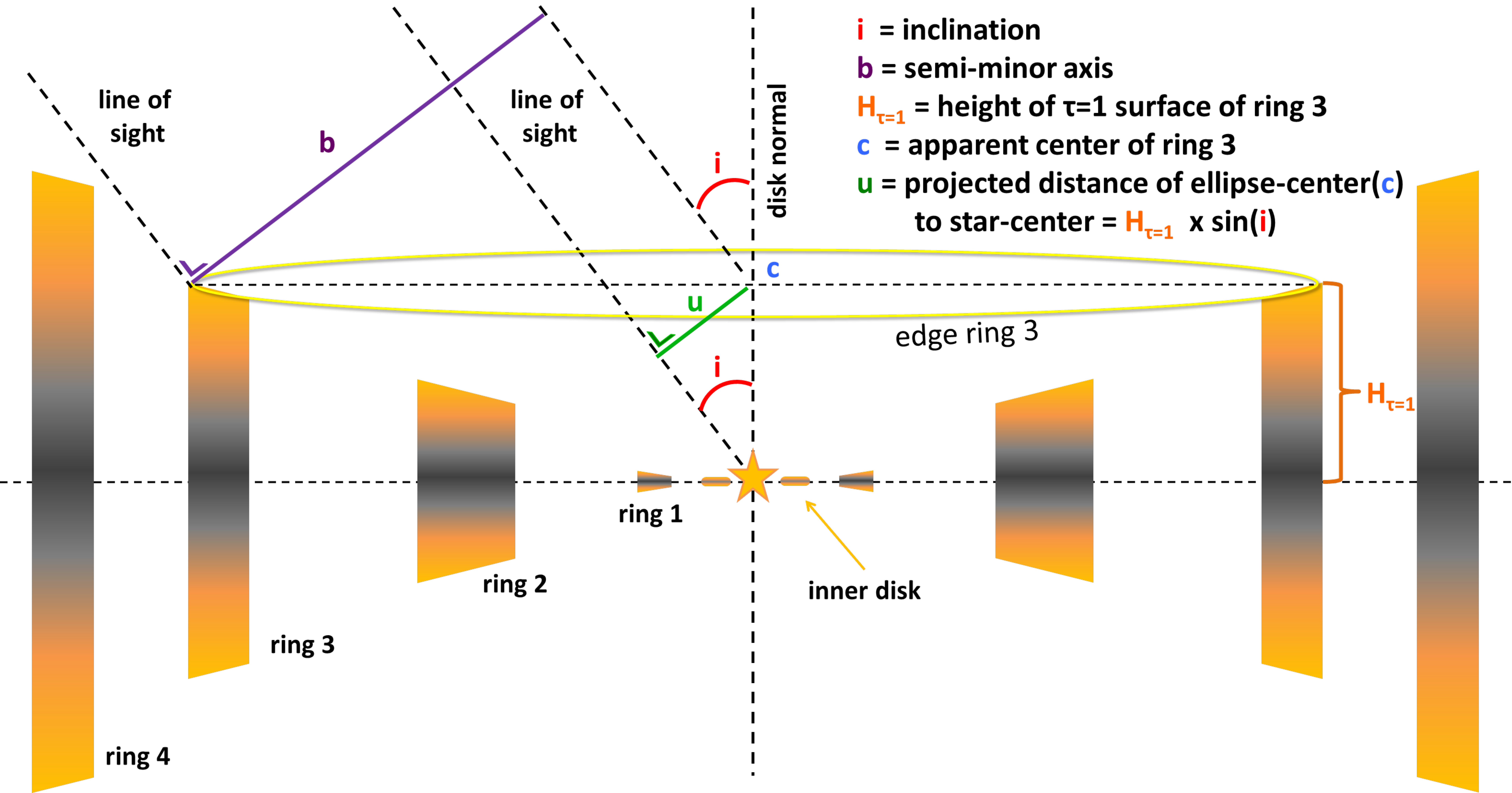}

\caption[]{Schematical cross section of the disk model suggested by our observations. The location and the height of the depicted rings is to scale, while the width of individual features is not. 
We use Ring\,3 as an example to show the geometrical relation between the height and center offset of individual rings assuming an axi-symmetric disk. Please note that the gaps in the disk are shown empty in this schematic view for simplicity, while in reality the gaps are likely not devoid of material.
This figure was adapted from de Boer et al., A\&A submitted.} 
\label{disk_schematic}
\end{figure*}

\begin{figure}
\centering
\includegraphics[scale=0.35]{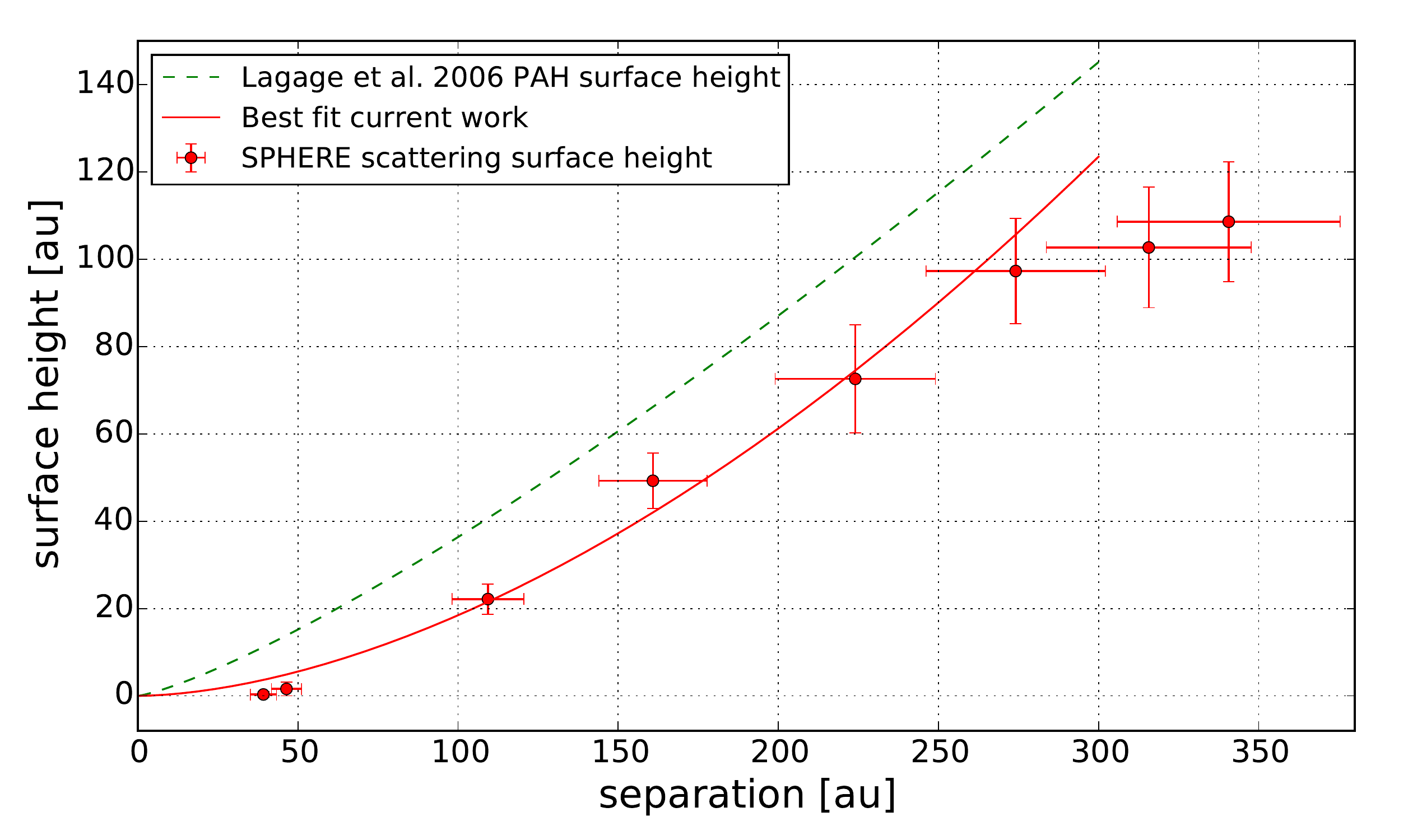}

\caption[]{Height of the scattering surface above the disk midplane as calculated from the center offset of the visible rings and gaps of the disk (red data points).
We added the height of the PAH emission surface calculated by \cite{2006Sci...314..621L} as dashed (green) line and our own best fit to the scattering surface height as solid (red) line.
The best fit to the scattered light data ignores the last two (furthest separated) data points, which deviate from a power law profile presumably due to a changing optical thickness of the disk.} 
\label{height}
\end{figure}

To quantify the geometry of the disk we fitted ellipses to the rings and gaps detected with high signal to noise in our SPHERE DPI and ADI data. 
Ellipses offset from the stellar position should describe the features sufficiently, assuming that we are looking at a axi-symmetric, inclined, flared disk.\\
Our fitting procedure consists of a Monte Carlo routine that created 10$^6$ elliptical annuli for each structure within boundaries such that the annuli trace the rings or gaps. 
We then measured the flux in each of these annuli and found the annulus which maximizes the flux in the case of the bright rings, or minimizes it in the case of the dark gaps. 
Since only the DPI data shows the full rings we used only this data set to determine the position angle and inclination of the disk. We specifically fitted Ring\,1 and Ring\,2 simultaneously in the unscaled image for this purpose.
To ensure a good fitting result the horizontal PSF artifact that is visible in the DPI image was masked. 
In addition, we only used the backscattering side of Ring\,2 to set initial boundaries for the fit, but masked it during the fitting procedure. 
This was done since the somewhat extended flux that we are detecting on the eastern side of Ring\,2 likely originates from the partially illuminated "wall" of Ring\,2, i.e. from a different height as the forward scattering western side of the ring.
This effect would be expected to be much less significant for the innermost ring, thus we included the full ring in the fit. The width of the measuring annulus was set to 2 pixels. 
This narrow width was chosen to avoid contaminating flux in the annulus from further inside the disk, which would dominate over the flux of the rings.\\
Using this procedure we find an inclination of the disk of 39.9$\pm$1.8\,deg, slightly smaller but consistent with the inclination of 42.8$^{+0.8}_{-2.5}$\,deg recovered by \cite{2006Sci...314..621L} with isophot fitting of their PAH emission image. 
We further recover a position angle of the major axis of the disk of 2.8$\pm$1.6\,deg. The uncertainties of these values were computed using the same procedure on the U$_\phi$ image for a noise estimation.
These results are also highly consistent with recent ALMA Cycle 0 observations of HD\,97048 (\citealt{2016arXiv160902011W}), who find a disk inclination of 41\,deg and a position angle of 3\,deg.\\
Using the measured inclination and position angle we then individually fitted Gap\,1 and Gap\,2 in the DPI image as well as Ring\,3 and Ring\,4 and the corresponding gaps in the ADI image. 
To constrain the size of the rings in the ADI image we set boundaries such that the rings should include the faint western backscattering region visible in the r$^2$ scaled DPI image.\\
We estimated reasonable error bars for each parameter by repeating the same fitting procedure with the same boundaries for each sub-structure in the U$_\phi$ image for Ring\,1, 2 and the associated gaps, and on the eastern side of the ADI image for Ring\,3 and 4.
This was done to determine the influence of the background noise on the fitting procedure. The resulting flux distributions after 10$^6$ measurements were fitted with a normal distribution. 
We then considered the width of these normal distributions for each sub-structure to determine a range of well fitting parameters. 
Finally we computed the standard deviation of each parameter within this range to estimate the uncertainty of each parameter. 
The results of our fitting procedure are listed in Tab.~\ref{tab: ellipses}. We also show the best fitting ellipses to the bright rings over-plotted in Fig~\ref{ellipse_fits}. \\
The increase of the offsets of the ellipses from the stellar position is most likely induced by the flaring of the disk. We used these values along with our recovered inclination to determine the surface height of the disk at which our scattered light signal originates. 
This was also done by \cite{2006Sci...314..621L} for their mid-IR observations. For an illustration of the involved geometry we refer to Fig.~\ref{disk_schematic} (adapted from de Boer et al, submitted to A\&A). We show our results in Fig.~\ref{height} and the corresponding Tab.~\ref{tab: height}. The height of the disk that we recover is slightly smaller than the one found by \cite{2006Sci...314..621L}. This is not surprising since they traced PAH molecule emission in their observation, which should originate at the disk surface where the disk is directly irradiated by stellar UV flux.
We find that the scattering surface height $H$ at a separation $r$ from the star can be described reasonably well with a single power law of the form $H(r) = 0.0064\,\mathrm{au} \cdot (r/1\,\mathrm{au})^{1.73} $ up to a separation of $\sim$270\,au. The last two data points at 316\,au and 341\,au deviate significantly from this power law.
This may be explained by the disk surface layer becoming optically thin at large separations such that the scattered light originates from a lower surface height. A similar behaviour is also visible at $\sim$350\,au for the PAH emission shown in Fig.~2 of \cite{2006Sci...314..621L}.\\
\cite{1997ApJ...490..368C} show that for an irradiated disk, a typical value for the flaring of 9/7\,$\approx$\,1.3 should be expected.  The exponent measured here is significantly larger than that theoretical value.  However, we have to keep in mind that the pressure scale height for which the 9/7 exponent has been derived is entirely dependent on the temperature and does not directly translate into the height where the disk becomes optically thick for photons flying radially, so where we would expect the scattering surface to lie.  One possible explanation for the stronger surface flaring would be, that HD97047 has a more shallow radial slope of the surface density that would push up the surface height in the outer regions.  Incidentally, this is consistent with the very low physical scale heigh that we measure in the inner disk region, expressed in our fit by the very low prefactor of the flaring powerlaw.  
Effectively, the region inside $\sim$50\,au is geometrically flat, pointing to a very low surface density or extremely effective dust settling.

\begin{table*}[t]
\small
 \centering
  \caption{Ellipse parameters fitted to the visible rings and gaps in our scattered light images. We give the offset of the ellipses from the central star position as well as the size of the major and minor axes for each fitted feature. We also give the size of the semi-major axis in au, since it directly corresponds to the physical radius of the de-projected rings. 
  Inclination and position angle were fitted simultaneously for all structures resulting in a single value.}
  \begin{tabular}{@{}lcccccccc@{}}
  \hline   
	\hline
 	 			& Gap\,1 	& Ring\,1 	& Gap\,2 	& Ring\,2	& Gap\,3	& Ring\,3 	& Gap\,4	& Ring\,4	\\
 	\hline
  $\Delta$ R.A. [mas]		& -1.3$\pm$2.7 	& -6.5$\pm$6.4 	& -90$\pm$11 	& -200$\pm$17 	& -295$\pm$39	& -395$\pm$24	& -417$\pm$33	& -441$\pm$29	\\
  $\Delta$ Dec [mas]		& 26.8$\pm$7.6	& 35.8$\pm$0.8 	& -9$\pm$16 	& -33$\pm$26 	& -12$\pm$32	& -28$\pm$15	& -44$\pm$29	& -64$\pm$25	\\
  semi-major axis [mas]		& 248.2$\pm$5.3	& 293.8$\pm$3.7	& 692$\pm$11	& 1018$\pm$28	& 1418$\pm$62	& 1735$\pm$24	& 1998$\pm$23	&2156$\pm$27	\\
  semi-major axis [au]		& 39.2$\pm$4.1	& 46.4$\pm$4.7	& 109$\pm$11	& 161$\pm$17	& 224$\pm$25	& 274$\pm$28	& 316$\pm$32	&341$\pm$35	\\
  semi-minor axis [mas]		& 190.4$\pm$4.1	& 225.4$\pm$2.4	& 530.9$\pm$8.1	& 781$\pm$15	& 1088$\pm$62	& 1331$\pm$22	& 1533$\pm$21	&1654$\pm$24	\\
  \hline
  inclination [deg]		& \multicolumn{8}{c}{39.9$\pm$1.8} 								\\
  PA [deg]			& \multicolumn{8}{c}{2.8$\pm$1.6} 									\\
 	
 \hline\end{tabular}

\label{tab: ellipses}
\end{table*}

\begin{table}
\small
 \centering
  \caption{Scattering surface height as function of stellar separation as calculated from the fitted ellipse offsets and inclination given in Tab.~\ref{tab: ellipses}.}
  \begin{tabular}{@{}lcc@{}}
  \hline   
	\hline
	&  Separation\,[au] & Height\,[au] \\
 	 \hline
Gap\,1	& 39.2$\pm$4.1		&	0.3$\pm$0.7	\\
Ring\,1	& 46.4$\pm$4.7		&	1.6$\pm$1.6	\\
Gap\,2	& 109.3$\pm$11.2		&	22.2$\pm$3.5	\\
Ring\,2	& 160.8$\pm$16.9		&	49.3$\pm$6.4	\\
Gap\,3 & 224$\pm$25			&		72.6$\pm$12.4	\\
Ring\,3	& 274.1$\pm$28.0		&	97.3$\pm$12.1	\\
Gap\,4 & 316$\pm$32				&	102.7$\pm$13.8		\\
Ring\,4	& 340.6$\pm$34.8		&	108.6$\pm$13.7	\\
 	\hline

 \hline\end{tabular}

\label{tab: height}
\end{table}

\subsection{Scattered light phase function}
\label{3.3}

\begin{figure}
\centering
\includegraphics[scale=0.48]{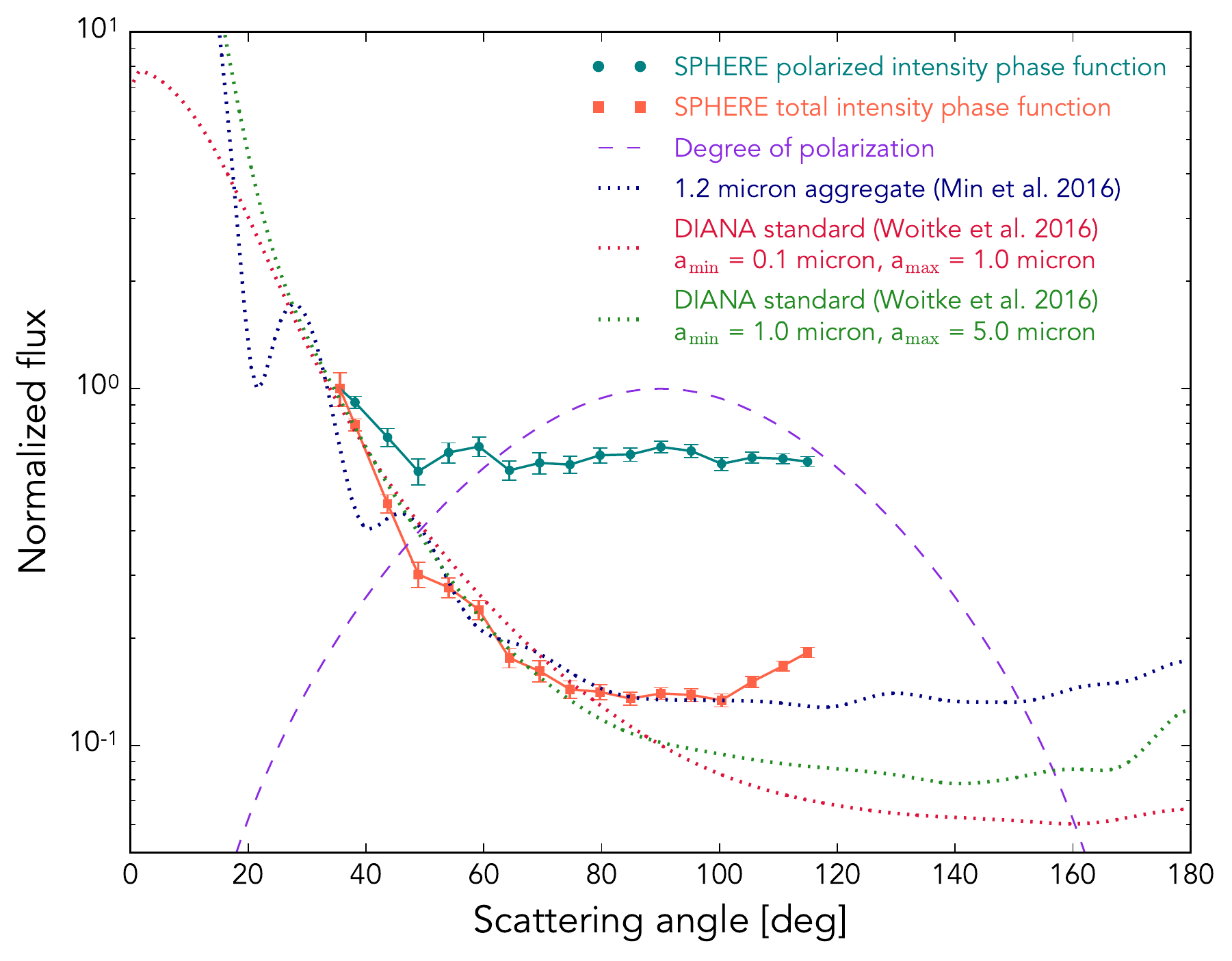}

\caption[]{Polarized intensity phase function (green data points) and reconstructed total intensity phase function (red data points) obtained from the SPHERE Q$_\phi$ image. The phase functions have been normalized to their peak value and the error bars show 1$\sigma$ uncertainties determined from the U$_\phi$ image. A bell-shaped degree of polarization (purple dashed line) was used to reconstruct the total intensity phase function. The dotted lines show numerically calculated phase functions in J-band for DIANA standard dust opacities \citep{woitke2016} and a compact dust aggregate \citep{min2016}.
} 
\label{phase}
\end{figure}

The surface brightness of an inclined protoplanetary disk will be affected by the phase function of the dust which depends on grain properties such as size, shape, structure and index of refraction \citep[e.g.][]{mishchenko2000}. Therefore, a measurement of the (polarization) phase function of the scattered light image will allow us to put constraints on the grain properties in the disk surface of HD\,97048. We follow the method from Stolker et al. (submitted to A\&A), which maps the scattered light image onto a power law shaped disk surface. In this way, we can calculate the disk radius and scattering angle in each pixel which is required to determine the scattering phase function and to construct $r^2$-scaled images that correct both for the inclination and thickness of the disk.\\
In Sect. \ref{3.2}, we used ellipse fitting to determine at several disk radii the height of the disk surface where in radial direction the optical depth is unity. The power law that was fitted to those data points (see Fig. \ref{height}) is used as input for the scattered light mapping from which we calculated $r^2$-scaled Q$_\phi$ images\footnote{We note that our scattered light surface height profile is likely too steep beyond $\sim$270\,au, thus the brightness increase in that region of the image might be somewhat overestimated.} (see Figs. \ref{hd97048-main} and \ref{ellipse_anno}) and determined the phase function of the SPHERE image along Ring 2 (130-160 au). The phase function calculation was not possible for Ring 1 because it shows strong artifacts from the telescope spider, neither for Ring 3 because the detection is of low S/N.\\
DPI observations measure polarized intensity so we have to correct the polarized intensity phase function for the degree of polarization in order to obtain a total intensity phase function. Scattering of light by protoplanetary dust grains gives rise to a scattering angle dependent degree of polarization which can often be approximated by a bell-shaped curve \citep[e.g.][]{mishchenko2002,min2005,murakawa2010}. Therefore, we use a Rayleigh polarization curve to reconstruct the total intensity phase function. Note that the peak value of the bell-shaped polarization curve does not affect the result because we work in normalized surface brightness units.\\
Figure \ref{phase} shows the polarized intensity phase function that is measured from the SPHERE Q$_\phi$ image, as well as the reconstructed total intensity phase function that is obtained from the ratio of the polarized intensity phase function and the degree of polarization. The polarized intensity phase function is close to isotropic for most of the scattering angles that are probed with the SPHERE DPI observations. The degree of polarization decreases towards forward and backward scattering angles which gives rise to the forward scattering peak in the reconstructed total intensity phase function since we probe scattering angles as small as $35\degr$. This result is supported by the ADI observation which shows that the near side of the disk is brighter than the far side in total intensity (see Fig. \ref{hd97048-main}).\\
Grains that are much smaller than the observed wavelength scatter light close to isotropic (Rayleigh regime) whereas grains that are comparable to or larger than the wavelength show a forward scattering peak and possibly a shallow backward scattering peak \citep[e.g.][]{bohren1983,min2005,min2016}. For comparison, we show in Fig. \ref{phase} two numerical phase functions that were calculated at 1.25 $\mu$m according to the DIANA standard dust opacities \citep{woitke2016} for a mixture of silicates and amorphous carbon grains that are porous (25\%) and irregularly shaped \citep[f$_{\mathrm{max}}=0.8$][]{min2005}. We chose two different grain size distributions, $0.1 < a < 1.0$ $\mu$m and $1.0 < a < 5.0$ $\mu$m, both with a power law exponent of -3.5. As expected, the forward scattering peak is stronger for the size distribution of larger grains. Although the phase functions seem to match quite well with the SPHERE data at smaller scattering angles, at intermediate scattering angles the phase functions deviate.\\
Additionally, Fig. \ref{phase} shows the phase function of a 1.2 $\mu$m compact dust aggregate, calculated by \citet{min2016} at 1.2 $\mu$m with the discrete dipole approximation (DDA). Note that this is a single grain size calculation so there are strong refraction features present in the phase curve. This phase function provides a better fit with the data because the turn over point (around $70\degr$) towards the isotropic part is present both in the SPHERE data and the aggregate calculation. Beyond $100\degr$, the reconstructed phase function starts to rise again and deviates from the aggregate phase function, but this could be an artifact of the horizontal brightness enhancement that is visible in the IRDIS DPI data. The aggregate structure of large dust grains ($2\pi a \gtrsim \lambda$ with $a$ the grains radius and $\lambda$ the photon wavelength) will provide the grains with aerodynamic support against settling from the HD\,97048 disk surface towards the midplane. We note that large grain aggregates are consistent with our earlier findings that the color of the disk appears grey between H and J band.

\section{Discussion of the scattered light gaps}

\begin{figure}
\centering
\includegraphics[scale=0.435]{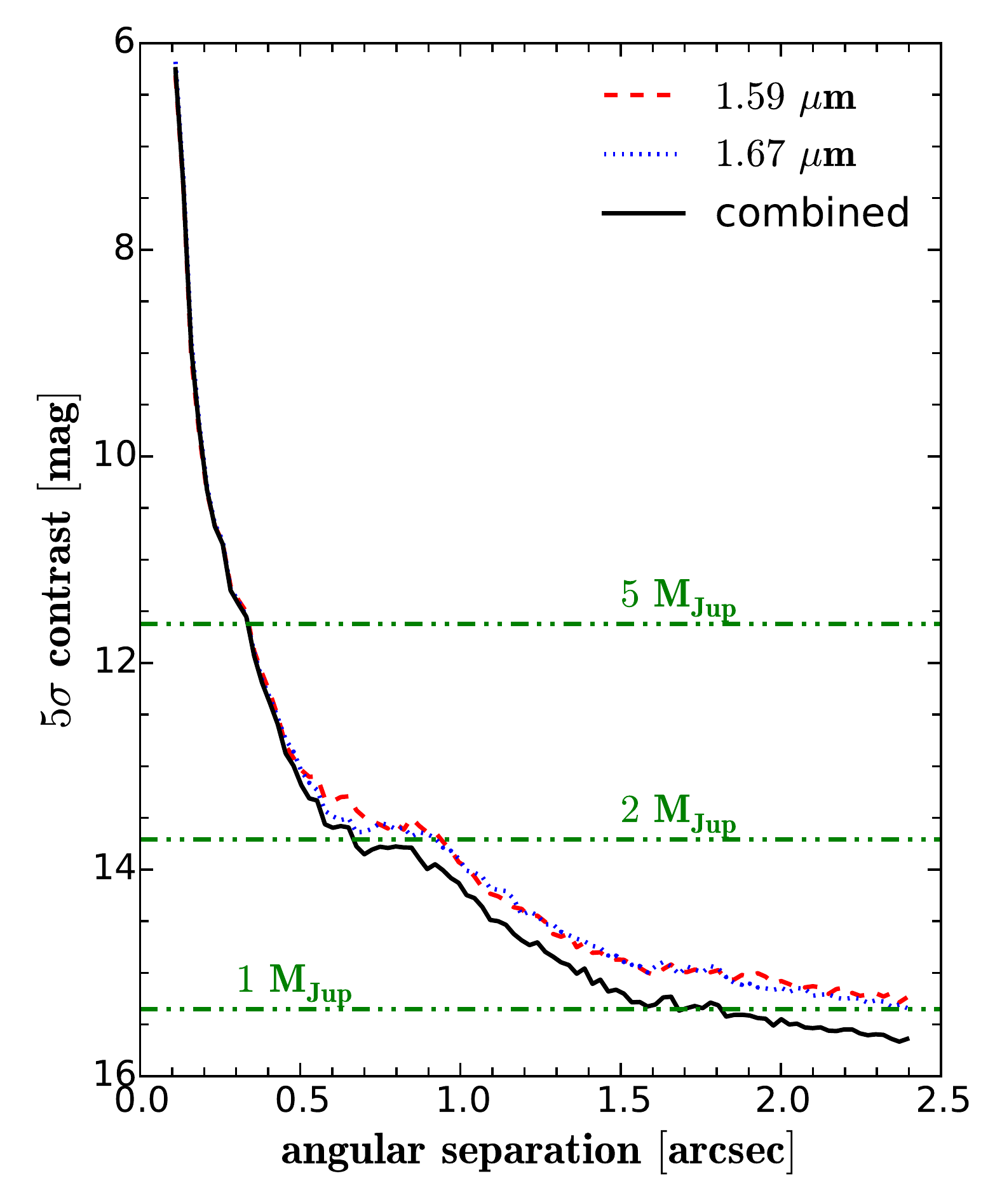}

\caption[]{Azimuthally averaged detection limits for the SPHERE ADI observation reduced with the TLOCI algorithm. We give the 5\,$\sigma$ detection limit in magnitudes. Assuming an age of the system of 2.5\,Myr and a distance of 158\,pc we calculated mass detection limits using the BT-SETTL model isochrones (green, dash-dotted lines).} 
\label{detect}
\end{figure}

Several explanations for the formation of rings and gaps have been proposed in protoplanetary disks, in particular for the recent mm-observations of HL\,Tau (\citealt{2015ApJ...808L...3A}) and TW\,Hya (\citealt{2016ApJ...820L..40A}), which show multiple ring structures as seen in the scattered light images of HD\,97048. 
Although the scattered light emission originates from the upper layers of the disk while the mm-emission comes from a deeper layer, the origin of the gaps might be directly connected.\\ 
For instance, planet-disk interaction can lead to gaps in the gas, small (micron-sized), and large grains (e.g. \citealt{2006MNRAS.373.1619R}, \citealt{2007A&A...474.1037F}, \citealt{2012A&A...545A..81P}, \citealt{2012ApJ...755....6Z}), although with different shape and location in small and large grains depending on the planet mass and disk viscosity (e.g. \citealt{2013A&A...560A.111D}, \citealt{2016MNRAS.459L..85D}). This segregation of small and large grains has been observed in different transition disks such as SR\,21 (\citealt{2013ApJ...767...10F}) and HD\,135344\,B (\citealt{2013A&A...560A.105G}). In the case of HD\,97048, we have four clear gaps at $\sim$0.25\,arcsec, $\sim$0.7\,arcsec, $\sim$1.4\,arcsec and $\sim$2.0\,arcsec distance from the star (along the semi-major axis). Analysis of the mm-emission in the u-v plane of this disk from ALMA-Cycle 0 observations (with a beam of ~0.9 x 0.52'') hint at the presence of two gaps at $\sim$0.6\,arcsec and $\sim$1.2\,arcsec (Fig 7., \citealt{2016arXiv160902011W}). 
The potential close match between the location and shape of the Gaps\,2 and 3 by ALMA and SPHERE suggests that if planets are responsible for the seen gaps, the mass of the embedded planets should be small ($\lesssim$ 1\,M$_\mathrm{Jup}$, see Fig 8 from \citealt{2013A&A...560A.111D}). 
The reasoning in this case would be, that the shape of the gaps in small grains would be the same as in the gas because they are well coupled and follow the gas distribution. 
Large grains, which are highly affected by radial drift would be accumulated at the location of the pressure maximum, which can be further out than the location of the edge of the gap in gas (or small grains) depending on the planet mass.
However, we note that the study by \citealt{2013A&A...560A.111D} was restricted to a single planet carving a gap in the disk and to cases where the planets are massive enough to open distinct gaps in the gas. 
In our case we directly detect 4 gaps and rings. Assuming that all of these structures are indeed caused by nascent planets, it would be expected that the location of the gap or ring in mm emission as well as in scattered light is constrained not just by the planet in the gap, but also by neighbouring planets.
Thus it might well be that the specific geometry of the HD\,97048 system does not allow for offsets between the observed gaps in mm emission and in scattered light, even if planets with masses larger than 1\,M$_\mathrm{Jup}$ are present.\\
To further constrain possible planet masses we used the empirical formula derived very recently by \cite{2016arXiv160303853K} from two dimensional hydrodynamic simulations. 
The width and position of the gap depends in this case on the mass ratio between companion and host star, the aspect ratio of the disk at the gap position, and the viscosity of the disk given by the viscous parameter $\alpha$ as defined by \cite{1973A&A....24..337S}.
We fitted a Gaussian with a linear slope to the surface brightness profile at the gap positions along the major axis and used the FWHM of the Gaussian as a measure of the gap width. We find a width of 61.8$\pm$5.5\,mas (9.8\,au) for Gap\,1 and 133$\pm$11\,mas (21\,au) for Gap\,2 in the SPHERE DPI image. Since detailed forward modelling is required to account for the effects of the ADI data reduction on the disk features, we did not inlcude Gap\,3 and Gap\,4 in this analysis.\\
The aspect ratio of the disk at the gap position can be estimated from our ellipse fitting, considering that the pressure scale height is approximately a factor of 4 smaller than the scattering height of small micron-sized particles if we assume the dust and the gas are well mixed (\citealt{2001ApJ...547.1077C}). 
This assumption seems reasonable especially in the case of Gap\,2 for which we find a gap width comparable to the scattering surface height. Since the gap width should be significantly larger than the pressure scale height to open a stable gap, a large ratio between the scattering surface height and the pressure scale height would be expected.
We note that the calculations of \cite{2016arXiv160303853K} were done for the gas in the disk, while we are tracing micron-sized dust particles. However, as is also noted in \cite{2016arXiv160303853K}, small dust particles are coupled well to the gas, and thus we would expect similar gap widths in both cases.
Assuming a range of disk viscosities between 10$^{-2}$ and 10$^{-4}$ we find necessary planet masses between 2.6\,M$_{\oplus}$ and 0.3M$_{\oplus}$ to carve out Gap\,1 and 0.7\,M$_{\mathrm{Jup}}$ to 0.1M$_{\mathrm{Jup}}$ for Gap\,2 (more massive planets for a higher viscosity).\\ 
This result would be consistent with our earlier estimate that planet masses should be below 1M$_{\mathrm{Jup}}$ based on the coincidence of the SPHERE and ALMA gaps (Gaps\,2 and 3).
The very small planet mass neccessary to open Gap\,1 might be plausible given that we find that the disk is basically flat at this small separation. Gap\,1 is not detected in the analysis of the ALMA data by \cite{2016arXiv160902011W}. 
This might be related to the lower resolution of their observations, but could be also be due to the low mass of a potentially responsible planet, which is then not expected to affect mm-sized grains (\citealt{2016MNRAS.459.2790R}).
However, we note that the smallest planet mass studied by \cite{2016arXiv160303853K} is 0.1\,M$_{\mathrm{Jup}}$ and thus our calculation for Gap\,1 represents an extrapolation from their results.
Furthermore, the recent study by \cite{2016MNRAS.459L...1D} using 3D smooted-particle hydrodynamics suggests that planets with masses smaller than 0.5\,M$_{\mathrm{Jup}}$ should not open gaps in gas and micron-sized dust particles. We thus caution that the mass that we find for a potential planet carving Gap\,1 should only be treated as a lower limit.\\
In any case the small planet masses we find for Gap\,1 and 2 are consistent with the non-detection of any close point sources in the SPHERE ADI data set. In Fig.~\ref{detect} we show the 5\,$\sigma$ contrast limits computed from this data set. 
The contrast is derived from the TLOCI image, in which the self-subtraction for each TLOCI area is estimated with injected fake planet signals. An azimuthal standard deviation is calculated to yield a 1-D radial contrast.
We use the BT-SETTL model isochrones (\citealt{2011ASPC..448...91A}) to convert the achievable magnitude contrast to limiting masses assuming a system age of 2.5\,Myr and a distance of 158\,pc as well as a stellar H-band magnitude of 6.67$\pm$0.05\,mag (\citealt{2003yCat.2246....0C}).
We find that in principle we would have detected planets with masses down to $\sim$2\,M$_{\mathrm{Jup}}$ beyond 1\,arcsec in the ADI data and planets with masses down to $\sim$5\,M$_{\mathrm{Jup}}$ as close as 0.3\,arcsec.
However, this does not take into account that the gaps that we are seeing are likely not completely devoid of material, and thus the thermal radiation of potential planets would be attenuated compared to the theoretical model values.\\
Other possible explanations for the origin of the rings and gaps include the effect of snow-lines in the dust density distribution (\citealt{2015ApJ...806L...7Z}; \citealt{2016ApJ...821...82O}) and pressure bumps with MHD origin (\citealt{2011ApJ...736...85U}, \citealt{2014ApJ...784...15S}, \citealt{2015A&A...574A..68F}). 
\cite{2015ApJ...815L..15B} recently showed the effect of the H$_2$O snow line on the distribution of dust particles by considering the changes of the aerodynamics of the dust grains when they loose their H$_2$O ice mantles. 
They showed that the H$_2$O snow line can have a significant effect on the distribution of small and large particles and decrease the dust density distribution of the small grains at the snow line location. 
However, for HD\,97048 the H$_2$O snow line is expected to be at 10\,mas (using the same model for the HD\,97048 disk as in \citealt{2013A&A...555A..64M}), which does not match any of the observed structures. 
Extending the H$_2$O snow line predictions for other relevant molecular lines, such as CO and ammonia (NH$_3$) for which freezing temperatures are ~20\,K and ~80\,K respectively (e.g. \citealt{2015ApJ...806L...7Z}), the expected disk structures should be located at 1.4\,arcsec and 0.3\,arcsec. These locations roughly coincide with Gap\,3 and Ring\,1 respectively.
For small disk viscosities ($\alpha$\,$\leq$\,10$^-3$) \cite{2015ApJ...815L..15B} found a rise of spectral index just outside the snow line which would produce a ring-like feature in mm-emission. This might be a possible explanation for Ring\,1 and Gap\,1, although we note that it is not clear if the same feature would be visible in scattered light.
High resolution ALMA observations in combination with detailed radiative transfer modelling might be able to determine if this is the case. For higher disk viscosities the spectral index of mm-emissions was found to drop at the position of the snowlines, which could be a potential explanation for Gap\,3. However, the same reservations apply regarding the visibility of such features in scattered light. 
Furthermore it is not clear if CO or NH$_3$ produce similarly strong effects as H$_2$O given their likely lower concentration in the disk.\\   
Alternatively, global pressure bumps predicted in MHD simulations can create multiple ring structures at mm-emission (e.g. \citealt{2016arXiv160305179R}). However, the structures expected in scattered light can be different since constant fragmentation of the mm-grains inside the bumps together with turbulent motion of the particles can lead to variations of the distribution of the small particles inside and outside these bumps (\citealt{2012A&A...538A.114P}). Currently there are also no scattered light predictions of such a scenario and more effort is required to show what shapes of gaps and rings at short and long wavelengths are expected in the MHD case. 
However, both possibilities (snow lines and/or MHD effects) cannot be ruled out for the origin of the observed rings and gaps in HD97048.

\section{Conclusions}

We observed the large circumstellar disk around HD\,97048 with SPHERE/IRDIS in scattered light using polarimetric and angular differential imaging and uncovered for the first time directly 4 gaps and rings in the disk. 
Our observations show very well the complementary nature of DPI and ADI observations. DPI observations enabled us to study the disk at the smallest angular separations and without self-subtraction effects, while the ADI observations provide deeper detection limits further out from the star.\\
The inclination and position angle of the disk that we extract from our observational data are consistent with the measurements by \cite{2006Sci...314..621L} as well as very recent ALMA Cycle 0 observations by \cite{2016arXiv160902011W}. Our innermost gap at $\sim$248\,mas (39\,au) and the following bright ring at $\sim$294\,mas (46\,au) are identical with the gap and ring at 34$\pm$4\,au inferred by \cite{2013A&A...555A..64M} from unresolved photometry and resolved Q-band imaging.\\
Our measurements suggest that the disk is flaring (assuming an axisymmetric disk) and that the scattering surface height profile of the disk is not significantly influenced by the observed structures. Specifically we find that the surface height can be described by a single power law up to a separation of $\sim$270\,au. Using this power law we derive the polarized scattering phase function of the disk.
Assuming a bell-shaped dependency of the degree of polarization of the scattering angle we find that the total intensity scattering phase function shows a turn-off at $\sim$70$^\circ$, consistent with 1.2 $\mu$m compact dust aggregates as calculated by \citet{min2016}.\\
We find that nascent planets are one possible explanation for the structures that we are observing. However, given the low planet masses needed to carve out the gaps that we detected, it is unlikely that the planet's thermal radiation is directly detectable by current generation planet search instruments such as SPHERE or GPI. This conclusion is strengthened by the fact that the gaps are most likely not completely devoid of material and thus any thermal radiation from a planet inside the gap would be attenuated by the remaining dust.\\
A detailed radiative transfer study, which includes both our scattered light observations and high-resolution ALMA observations, is required for a more complete understanding of the disk structure and distribution of small and large grains in particular.
We note that high resolution ALMA observation by \cite{2016arXiv160902488V} have just become available at the time of publication of our observations, which will enable such a study.


\begin{acknowledgements}
SPHERE is an instrument designed and built by a consortium consisting of IPAG (Grenoble, France), MPIA (Heidelberg, Germany), LAM (Marseille, France), LESIA (Paris, France), Laboratoire Lagrange (Nice, France), INAF - Osservatorio di Padova (Italy), Observatoire de Geneve (Switzerland), ETH Zurich (Switzerland), NOVA (Netherlands), ONERA (France) and ASTRON (Netherlands) in collaboration with ESO. SPHERE was funded by ESO, with additional contributions from CNRS (France), MPIA (Germany), INAF (Italy), FINES (Switzerland) and NOVA (Netherlands). SPHERE also received funding from the European Commission Sixth and Seventh Framework Programmes as part of the Optical Infrared Coordination Network for Astronomy (OPTICON) under grant number RII3-Ct-2004-001566 for FP6 (2004-2008), grant number 226604 for FP7 (2009-2012) and grant number 312430 for FP7 (2013-2016). 
We thank Catherine Walsh for sharing her results with us before publication. P. Pinilla is supported by a Royal Netherlands Academy of Arts and Sciences (KNAW) professor prize.
This research has made use of the SIMBAD database as well as the VizieR catalogue access tool, operated at CDS, Strasbourg, France. This research has made use of NASA's Astrophysics Data System Bibliographic Services. 
Finally CG would like to thank Donna Keeley for language editing of the manuscript.

\end{acknowledgements}

%
%

\bibliographystyle{aa} 
\bibliography{myBib} 

\end{document}